\documentclass{article}
\usepackage{spconf}

\usepackage[cmex10]{amsmath}
\usepackage{fixltx2e}
\usepackage{url}

\usepackage[T1]{fontenc}
\usepackage{amssymb,amsthm}
\usepackage{dsfont}
\usepackage{mathtools}
\usepackage{longtable}
\usepackage{booktabs}
\usepackage{tabularx}
\usepackage{ltxtable}
\usepackage{dcolumn}
\usepackage{lscape}
\usepackage{bm}

\newcommand{\R}{\mathds{R}}
\newcommand{\N}{\mathds{N}}

\newcommand{\Q}{\mathds{Q}}

\newcommand{\st}{{\rm s.t.}}
\newcommand{\norm}[1]{\lVert {#1} \rVert}
\newcommand{\abs}[1]{\lvert {#1} \rvert}
\newcommand{\suchthat}{\,:\,}
\newcommand{\define}{\coloneqq}

\newcommand{\NP}{\textsf{NP}}

\renewcommand{\P}{\textsf{P}}

\newcommand{\A}{\bm{A}}
\newcommand{\x}{\bm{x}}
\renewcommand{\b}{\bm{b}}
\newcommand{\Omeg}{\bm{\Omega}}
\newcommand{\omeg}{\bm{\omega}}
\newcommand{\z}{\bm{z}}
\newcommand{\e}{\bm{e}}

\newtheorem{theorem}{Theorem}

\newtheorem{corollary}{Corollary}
\newtheorem{remark}{Remark}

\begin{document}
%
\title{PROJECTION ONTO THE COSPARSE SET IS NP-HARD}
%
%
%

\name{Andreas M. Tillmann$\hspace{1pt}^\ddagger$\sthanks{Dolivostr. 15, 64293 Darmstadt, Germany. A.T. acknowledges funding by the Deutsche Forschungsgemeinschaft (DFG) within the project ``Sparse Exact and Approximate Recovery'' under grant PF 709/1-1.} \qquad R\'{e}mi Gribonval$\hspace{1pt}^\circ$\sthanks{Rennes-Bretagne Atlantique Research Center, Campus de Beaulieu, 35042 Rennes, France. R.G. acknowledges funding by the European Research Council within the PLEASE project under grant ERC-StG-2011-277906.\newline \hspace*{1.5em} This work has been accepted for publication by the IEEE. Copyright will be transferred without notice, after which this version may no longer be accessible.} \qquad Marc E. Pfetsch$\hspace{1pt}^\ddagger$}
\address{$^\ddagger$ Research Group Optimization, TU Darmstadt$^*$\\
  $^\circ$ Inria$^\dagger$}

\maketitle

\begin{abstract}
  The computational complexity of a problem arising in the context of
  sparse optimization is considered, namely, the projection onto the
  set of $k$-cosparse vectors w.r.t. some given matrix
  $\Omeg$. 
  It is shown that this projection problem is (strongly) \NP-hard,
  even in the special cases in which the matrix $\Omeg$ contains only
  ternary or bipolar coefficients. Interestingly, this is in contrast
  to the projection onto the set of $k$-sparse vectors, which is
  trivially solved by keeping only the $k$ largest coefficients.
\end{abstract}


%

\section{Introduction}\label{sect:intro}

A central problem in compressed sensing (CS), see, e.g.,
\cite{D06,E10,FR11}, is the task of finding a sparsest solution to an
underdetermined linear system, i.e.,
\begin{equation}\label{prob:P0}
  \min\,\norm{\x}_0\quad\st\quad \A\x=\b,\tag{${\rm P}_0$}
\end{equation}
for a given matrix $\A \in \R^{m \times n}$ with $m < n$ and right
hand side vector $\b\in\R^m$, where~$\norm{\x}_0$ denotes the
$\ell_0$-quasi-norm, i.e., the number of nonzero entries in~$\x$. This
problem is known to be strongly \NP-hard, cf.~[MP5] in~\cite{GJ79};
the same is true for the variant in which $\A\x=\b$ is replaced by
$\norm{\A\x-\b}_2\leq\varepsilon$, see~\cite{N95,DMA97}.

Two related problems arise in signal and image processing, where the
unknown signal $\x$ to be estimated from a low-dimensional observation
$\b=\A\x$ cannot directly be modeled as being sparse. In the most
standard approach, $\x$ is assumed to be built from the superposition
of few building blocks or \emph{atoms} from an overcomplete dictionary
${\bm D}$, i.e., $\x={\bm D}\z$, where the representation vector $\z$
is sparse. The problem of minimizing $\norm{\z}_0$ such that $\A{\bm
  D}\z=\b$ is obviously also \NP-hard.

The alternative \emph{cosparse analysis model} \cite{NDEG11} assumes
that $\Omeg\x$ has many zeros, where $\Omeg$ is an analysis
operator. Typical examples include finite difference operators: They
are closely connected to total variation minimization and defined as
computing the difference between adjacent sample values (for a signal)
or pixel values (for an image). The cosparse optimization problem of
interest reads
\begin{equation}\label{prob:CP0}
  \min\,\norm{\Omeg\x}_0\quad\st\quad \A\x=\b,\tag{${\rm C}_0$}
\end{equation}
and was also shown to be \NP-hard \cite[Section~4.1]{NDEG11}.

Due to the hardness of problem~\eqref{prob:P0}, various heuristics
have been developed. A popular approach is well-illustrated by the
Iterative Hard Thresholding (IHT) algorithm, which iterates a gradient
descent step to decrease the error $\norm{\A\tilde{\x}-\b}_2$ and a
hard-thresholding step, i.e., the (Euclidean) projection onto the set
of $k$-sparse vectors. Under restricted isometry properties (RIP) on
$\A$, the IHT algorithm can be proven to converge to the solution
of~\eqref{prob:P0} \cite{BD09,BD10}. A desirable RIP often holds with
high probability when $\A$ has i.i.d. sub-Gaussian entries, see, e.g.,
\cite{BDDW08}, but is (strongly) \NP-hard to verify in
general~\cite{PT12}.

Adaptations of IHT and related algorithms to the cosparse analysis
setting have been proposed \cite{GNGD11,GNEGD12}, based on a general
scheme for unions of subspaces \cite{B11}. A key step is the
projection onto the set of $k$-cosparse vectors, as an analogous
replacement for hard-thresholding. 

The main contribution of this note is to show that this projection is
in fact strongly \NP-hard in general, which contrasts with the
extremely simple and fast hard-thresholding operation.

\section{Complexity of Cosparse Projection Problems}\label{sect:result}

This section is mainly devoted to proving the following central result.
\begin{theorem}\label{thm:ProjNP}
  Consider any $p\in\N\cup\{\infty\}$, $p>1$, and let $q=p$ if
  $p<\infty$ and $q=1$ if $p=\infty$.  Given $\Omeg\in\Q^{r\times n}$
  ($r>n$), $\omeg\in\Q^n$, and a positive integer $k\in\N$, it is
  \NP-hard in the strong sense to solve the $k$-cosparse $\ell_p$-norm
  projection problem 
  \begin{equation}\label{prob:ProjOpt}
    \min_{\z\in\R^n}\,\{\,\norm{\omeg-\z}_{p}^{q}\suchthat\norm{\Omeg \z}_0\leq k\,\}, \tag{$k$-CoSP$_p$}
  \end{equation}
  even when $\omeg\in\{0,1\}^n$ (with exactly one
  nonzero entry) and $\Omeg$ contains only entries in $\{-1,0,+1\}$ or
  $\{-1,1\}$, respectively.
\end{theorem}

Let us first provide some comments on the statement of the theorem.

\begin{remark}\label{rem:approximation}
  Theorem~\ref{thm:ProjNP} shows the \NP-hardness of~\eqref{prob:ProjOpt}
  in the strong sense (i.e., \eqref{prob:ProjOpt} is strongly
  NP-hard). Thus, since the objective is always polynomially bounded
  (never exceeding $\norm{\omeg}_p^p$ or $\norm{\omeg}_\infty$), not even a
  \emph{fully polynomial-time approximation scheme (FPTAS)} can exist,
  unless \P$=$\NP. An FPTAS is an algorithm that, given an approximation
  factor $\epsilon>0$, computes a solution that has an objective function
  value at most $(1+\epsilon)$ times the minimum value, with running time
  polynomial in the encoding length of the instance and
  $1/\epsilon$. Moreover, there also cannot exist a \emph{pseudo-polynomial
    (exact) algorithm}, i.e., one that runs in time polynomial in the
  \emph{numeric} value of the input, but not its \emph{encoding length},
  unless \P$=$\NP; for details, see~\cite{GJ79}, and also \cite{KV08}.
\end{remark}

\begin{remark}
  It is perhaps not immediately clear why the ``$\min$''
  in~\eqref{prob:ProjOpt} is attained, since the constraint set is
  not bounded, in general. Nevertheless, the infimum is finite since
  the objective is lower-bounded by zero.
  Moreover, since $\z=0$ is always feasible, the optimal value is
  upper-bounded by $\norm{\omeg}_p^q$. Since the 
  level set $\{\,\z\suchthat\norm{\Omeg\z}_0\leq
  k,~0\leq\norm{\omeg-\z}_p^q\leq\norm{\omeg}_p^q\,\}$ is compact, the
  infimum is attained, justifying the use of ``$\min$'' instead of
  ``$\inf$'' in~\eqref{prob:ProjOpt}. (See also
  Remark~\ref{rem:NPcompleteness} below.)
\end{remark}

Before stating the proof of Theorem~\ref{thm:ProjNP}, we need some
preliminaries. Let \textsc{MinULR}$_0^=$($\A$,$K$) be the problem to
decide, for a given matrix $\A\in\Q^{m\times n}$ and a positive integer
$K\in\N$, whether there exists a vector $\z\in\R^n$ such that $\z\neq
0$ and at most $K$ of the $m$ equalities in the system $\A\z=0$ are
violated (\textsc{MinULR} stands for ``minimum number
of unsatisfied linear relations''). This homogeneous
equality version, in which the trivial all-zero solution is
excluded, was proven to be \NP-complete~\cite{AK95}, and even
\NP-hard to approximate within any constant factor~\cite{ABSS93}; more
results about the problem's approximation complexity can be found in
\cite{AK98}. For the proof of Theorem~\ref{thm:ProjNP}, we utilize the
following result:
\begin{theorem}[Theorem~1 and Corollary~2 in \cite{AK95}]\label{thm:MinULRternaryNP}\label{cor:MinULRbipolarNP}
  The problem \textsc{MinULR}$_0^=$($\A$,$K$) is strongly \NP-hard, even when $\A$
  contains only ternary entries, i.e., $\A\in\{-1,0,+1\}^{m\times n}$, or for bipolar $\A\in\{-1,+1\}^{m\times n}$.
\end{theorem}

The cited results from \cite{AK95} are in fact more general than the
statements we use here in that they also hold for inhomogeneous
systems, where the right hand side vector is part of the
input. Moreover, \cite{AK95} deals with the complementary problem of
\textsc{MinULR}$_0^=$, namely the \emph{Maximum Feasible Subsystem}
(\textsc{MaxFS}) problem (w.r.t. homogeneous linear equality systems)
in which one asks for the largest possible number of simultaneously
satisfied equalities from a given linear system. It is easy to see
that \textsc{MinULR} and \textsc{MaxFS} are equivalent when solved to
optimality (see also \cite{AK98}).

Note also that the above two results state \NP-hardness in the strong
sense, although this is not made explicit in the original
version~\cite{AK95}. The \NP-hardness proofs, however, are by
reduction from the \emph{Exact Cover by 3-Sets} (X3C) problem, which
is well-known to be strongly \NP-hard (see, e.g., \cite{GJ79}), and
preserve polynomial boundedness of the constructed numbers as well as
of their encoding lengths.

\begin{proof}[\bf\itshape Proof of Theorem~\ref{thm:ProjNP}]
  Let $p \in \N \cup \{\infty\}$, $p>1$, and $q=p$ if $p<\infty$ and $q=1$
  if $p=\infty$. Given an instance $(\A,K)$ of \textsc{MinULR}$_0^=$ (w.l.o.g., 
  $\A\in\Q^{r\times n}$ with $r>n$), we will reduce it to~$n$ instances
  of~\eqref{prob:ProjOpt}.

  For all $i=1,\dots,n$, we define a $k$-cosparse projection instance
  given by $\Omeg =\A$, $\omeg =\e_i$ and $k=K$ (where $\e_i$ denotes
  the $i$-th unit vector in $\Q^n$). Note that each such instance
  obviously has encoding length polynomially bounded by that of $\A$
  and~$K$. Writing
  \begin{align*}
    &f^{(p)}_{i,k}\define \min_{\z\in\R^n}\{\,\norm{\e_i-\z}_p^{q}\suchthat\norm{\Omeg \z}_0 \leq k\,\}\\
    \text{ and }\quad &f^{(p)}_k \define \min_{1\leq i\leq n}\,f^{(p)}_{i,k},
  \end{align*}
  we observe that since $\z=0$ is always a feasible point, it holds
  that $f^{(p)}_{i,k} \leq 1$ for all $i$, and hence $f^{(p)}_{k} \leq
  1$.  We claim that $f^{(p)}_k < 1$ if and only if there exists a
  nonzero vector $\z$ that violates at most~$k = K$ equations in $\A\z
  = 0$.
  \begin{enumerate}
  \item If $f^{(p)}_k < 1$ (i.e., there exists some
    $i\in\{1,\dots,n\}$ such that $f^{(p)}_{i,k}<1$) then there exists
    a vector $\z$ such that $\norm{\Omeg \z}_0 \leq k$ and
    $\norm{\e_{i}-\z}_{p}^{q} < 1$ (in particular, $\z_i\neq 0$). But
    this of course means that at most $K$ equalities in the system
    $\A\z=0$ are violated, i.e., \textsc{MinULR}$_0^=(\A,K)$ has a
    positive answer.
  \item Conversely, assume that there exists a nonzero vector~$\z$
    that violates at most~$K$ equations in $\A\z = 0$, i.e.,
    $\norm{\Omeg\z}_0 \leq k$ and \textsc{MinULR}$_0^=(\A,K)$ has a
    positive answer. We will prove that 
    \begin{equation}
    f^{(p)}_{k} \leq \min_{i,\lambda} \|\e_{i}-\lambda \z\|_{p}^{q} < 1.
    \end{equation}
    In fact, for an arbitrary scalar $\lambda\in\R$, $\z'\define
    \lambda\z$ obeys $\norm{\Omeg\z'}_0\leq k$ as well.  If $\z$
    contains only one nonzero component~$z_i$, $\lambda$ can be
    chosen sucht that $\z' = \e_i$ and consequently
    $f^{(p)}_k=f^{(p)}_{i,k} = 0$, which shows the claim. Thus, it 
    remains to deal with the case in which $\z$ contains at least two
    nonzero components. Consider some~$i$ such that $z_i \neq
    0$. Without loss of generality, we can also assume that $z_{i} >
    0$ (otherwise replace $\z$ with $-\z$). We consider the case
    $p < \infty$ first. It holds that
    \begin{align*}
      \norm{\e_i - \lambda\z}_p^p ~=~ &\norm{\lambda\z}_p^p - \abs{\lambda z_i}^p + \abs{1-\lambda z_i}^p\\
      =~ &\abs{\lambda}^p \big(\norm{\z}_p^p - \abs{z_i}^p\big) + \abs{1-\lambda z_i}^p
    \end{align*}
    for any $\lambda$. The function $\abs{1-\lambda z_i}^p$ is convex (in
    $\lambda$), and it is easy to see that we have $\abs{1-\lambda z_i}^p
    \leq 1 - \lambda z_i$ for $0 \leq \lambda \leq
    1/z_i$. Consequently, for $0 \leq \lambda \leq 1/z_i$,
    \begin{align}
      \nonumber &\abs{\lambda}^p \big(\norm{\z}_p^p - \abs{z_i}^p\big) + \abs{1-\lambda z_i}^p \\
      \label{eq:dist_bound}\leq~ &1 + \abs{\lambda}^p \big(\norm{\z}_p^p - \abs{z_i}^p\big) - \lambda z_i.
    \end{align}
    Since $\z$ contains at least two nonzero components, it follows
    that $\alpha \define \norm{\z}_p^p - \abs{z_i}^p > 0$. Since
    $p>1$, it is easy to see that $\abs{\lambda}^p \alpha - \lambda
    z_i < 0$ for sufficiently small positive $\lambda$. In
    conclusion, there exists a solution~$\lambda \z$ with $\norm{\e_i
      - \lambda\z}_p^p < 1$ (and thus $f^{(p)}_k < 1$), which proves
    the reverse direction for $p < \infty$.

    Finally, for the case $p=\infty$, note that
    \[
    \min_{i,\lambda}\norm{\e_i-\lambda\z}_\infty\leq\min_{i,\lambda}\norm{\e_i-\lambda\z}_{\tilde{p}},
    \]
    for all 
    $1<\tilde{p}<\infty$, 
    whence we can employ the above reasoning to reach the same
    conclusion.
  \end{enumerate}
  To summarize, we could reduce the \NP-hard problem
  \textsc{MinULR}$_0^=(\A,K)$ to $n$ instances of the $k$-cosparse
  projection problem~\eqref{prob:ProjOpt}, which therefore is \NP-hard
  as well. Moreover, by Theorem~\ref{thm:MinULRternaryNP} (and since
  all numerical values in the constructed instances obviously remain
  polynomially bounded by the input size), the \NP-hardness
  of~\eqref{prob:ProjOpt} holds in the \emph{strong} sense, even for
  $\Omeg$ $(=\hspace{-0.3em}\A)$ with ternary or bipolar coefficients
  and, by the construction above, for $\omeg$ binary (and indeed, with
  only one nonzero entry).
\end{proof}

\begin{remark}\label{rem:proofNotFor_pEQ1}
  The reduction in the proof of Theorem~\ref{thm:ProjNP} does not work
  for $0 < p \leq 1$.
  Indeed, consider the vector with all entries equal to $1$, $\z =
  \mathds{1}$.  It is not hard to verify that
  $\norm{\e_i-\lambda\z}_p^p =
  \abs{1-\lambda}^p+(n-1)\abs{\lambda}^p\geq 1$ for all~$i$
  and~$\lambda$. Thus, $\min_{i,\lambda}\norm{\e_i-\lambda\z}_p^p\geq
  1$ does not imply $\z=0$ when $p\in (0,1]$; vice versa, $\z \neq 0$
  does not imply that $\min_{i,\lambda}\norm{\e_i-\lambda\z}_p^p < 1$.
\end{remark}

\begin{remark}
  It is noteworthy that the claim in the proof of
  Theorem~\ref{thm:ProjNP} is in fact true for every real-valued
  $p>1$. However, one may encounter irrational (i.e., not finitely
  representable) numbers when working with $p$-th powers for arbitrary
  real, or even rational, $p$. Since retaining finite rational
  representations is crucial for \NP-hardness proofs, we restricted
  ourselves to integer~$p$ (or $p=\infty$).
\end{remark}

The following result is an immediate consequence of Theorem~\ref{thm:ProjNP}.
\begin{corollary}\label{cor:ProjArgNP}
  It is strongly \NP-hard to compute a minimizer for the
  problem~\eqref{prob:ProjOpt}, even under the input data restrictions
  specified in Theorem~\ref{thm:ProjNP}. In particular, it is strongly
  \NP-hard to compute the Euclidean projection
  \begin{equation}\label{prob:ProjOptArg}
    \Pi_{\Omeg,k}(\omeg)\define\arg\min_{\z\in\R^n}\,\{\,\norm{\omeg-\z}_{2}\suchthat\norm{\Omeg \z}_0\leq k\,\}.\tag{$k$-CoSP}
  \end{equation}
\end{corollary}
\begin{proof}
  Clearly, if a minimizer was known, we would also know the optimal
  value of~\eqref{prob:ProjOpt}. Hence, computing a minimizer is at
  least as hard as solving~\eqref{prob:ProjOpt}, and the complexity
  results of Theorem~\ref{thm:ProjNP} carry over directly. In
  particular, computing $\Pi_{\Omeg,k}(\omeg)$ is also strongly
  \NP-hard, since Theorem~\ref{thm:ProjNP} applies to the
  $\ell_2$-norm ($p=2$) and the minimizers of $\norm{\omeg-\z}_2$ (the
  objective in~\eqref{prob:ProjOptArg}) are of course the same as
  those of $\norm{\omeg-\z}_2^2$.
\end{proof}

Let us comment on a few more subtle aspects
concerning~\eqref{prob:ProjOpt} and the proof of
Theorem~\ref{thm:ProjNP}.
\begin{remark}
  The above reduction from \textsc{MinULR}$_0^=$ is an example of what
  is called a \emph{Turing reduction}; more commonly used are
  \emph{Karp reductions}, cf.~\cite{GJ79,KV08}. In the latter, the
  known \NP-hard problem is reduced to a \emph{single instance} of the
  problem under consideration. For~\eqref{prob:ProjOpt} with $p \in
  \N, p>1$ (excluding the case $p=\infty$), we could also obtain a
  Karp reduction by constructing the instance
  \begin{align*}
  \tilde{\Omeg}^\top &\define (\A^\top,\dots,\A^\top)\in\Q^{n\times rn},\\
  \tilde{\omeg}^\top &\define(\e_1^\top,\dots,\e_m^\top)\in\{0,1\}^{n^2},\quad \tilde{k} \define nK,
  \end{align*}
  which is obviously still polynomially related to the input $(\A, K)$
  of the given \textsc{MinULR}$_0^=$ instance. Then, defining
  \[
  \tilde{f}^{(p)}_{\tilde{k}} \define
  \min_{\z\in\R^{rn}}\{\,\norm{\tilde{\omeg} - \z}_p^{p} \suchthat \norm{\tilde{\Omeg}
    \z}_0 \leq \tilde{k}\,\},
  \]
  it is easy to see that $\tilde{f}^{(p)}_{\tilde{k}} <
  n\,(=\norm{\tilde{\omeg}}_p^p)$ if and only if
  \textsc{MinULR}$_0^=(A,K)$ has a positive answer. In particular, a
  solution can be seen to be comprised of the solutions to the $n$
  separate \eqref{prob:ProjOpt}-problems considered in the Turing
  reduction, stacked on top of each other. We chose the Turing
  reduction for our proof of Theorem~\ref{thm:ProjNP}
  because 
  it allows us to conclude \NP-hardness even if $\omeg$ is a unit
  vector, i.e., binary with exactly one nonzero entry, and also for
  $p=\infty$.
\end{remark}
\begin{remark}\label{rem:NPcompleteness}
 The optimization problem~\eqref{prob:ProjOpt} could be replaced by 
 its corresponding decision version:
 \begin{align}
   \nonumber &\text{\rm Does there exist some }\z\in\R^n\text{\rm such that }\\
   \label{prob:ProjOptDec}&\norm{\omeg-\z}_p^q <\gamma~\text{\rm and }\norm{\Omeg\z}_0\leq k\,?\tag{$k$-CoSP$_p$-Dec}
 \end{align}
 Indeed, the proof of Theorem~\ref{thm:ProjNP} corresponds to showing
 hardness of this decision problem for $\gamma=1$; the Karp reduction
 sketched in the previous remark would use $\gamma=n$.

 Note that~\eqref{prob:ProjOptDec} is in fact contained in {\NP} (at
 least for $p=1$, $2$, or $\infty$): 
 The set
 $\{\,\z\suchthat\norm{\Omeg\z}_0\leq k\,\}$ defined by the constraints yields (exponentially
 many) affine subspaces of $\R^n$ defined by $n-k$ or more homogeneous
 equalities, and the projection of some $\omeg$ onto it
 (w.r.t. $\norm{\cdot}_p^q$) is clearly equivalent to that onto one
 (unknown) of these affine subspaces. For $p=2$, the (Euclidean)
 projection onto such spaces has a known explicit formula which keeps
 all entries in the solution rational if the input ($\Omeg$) is
 rational. Similarly, for $p=1$ or $p=\infty$, \eqref{prob:ProjOpt}
 can be seen as a linear program over the unknown correct affine
 subspace; hence, here, the solution is also rational.

 Thus, for $p=1$, $2$, or $\infty$, a certificate of a positive answer
 exists that has encoding length polynomially bounded by that of the
 input. Hence, by Theorem~\ref{thm:ProjNP} and the above discussion,
 we have that \eqref{prob:ProjOptDec} is in fact \emph{strongly \NP-complete}
 for $p=2$ or $p=\infty$. (The case $p=1$ is not covered by
 Theorem~\ref{thm:ProjNP}, and for the remaining values of $p$ it is
 not immediately clear how one could guarantee the existence of a
 certificate with polynomially bounded encoding length.)
\end{remark}

\section{Conclusions}\label{sect:conclusions}

Theorem~\ref{thm:ProjNP} and Corollary~\ref{cor:ProjArgNP} show that
no polynomial algorithm to compute the projection onto the set of
$k$-cosparse vectors can exist unless \P$=$\NP.

In theoretical algorithmic applications of the Euclidean $k$-cosparse
projection operation \eqref{prob:ProjOptArg} in \cite{GNGD11,GNEGD12},
it had so far been assumed that the projection problem ($k$-CoSP$_2$)
can be approximated efficiently. While our result refutes this
assumption to a certain degree (cf. Remark~\ref{rem:approximation}),
it is not clear whether other (general polynomial-time) approximation
algorithms exist that may still be useful in practice despite
exhibiting theoretical running time bounds that depend exponentially
on (at least) the approximation quality~$1/\epsilon$.

Moreover, as for most NP-hard problems, there are specific instances
which are known to be much easier to handle than the general case. For
instance, when $\Omeg$ is the identity (or, more generally, a unitary)
matrix, hard-tresholding---i.e., zeroing all but the $k$ entries with
largest absolute values---achieves the projection onto the
$k$-cosparse set w.r.t. any $\ell_p$-norm. Other examples include the
Euclidean case ($p=2$) when $\Omeg$ is the 1D finite difference
operator or the 1D fused Lasso operator, respectively: The projection
is then achieved using dynamic programming \cite{GNEGD12}.

For the problem \textsc{MinULR}$_0^=$ (or its minimization variant,
respectively), strong non-approximability results were derived in
\cite{ABSS93,AK95,AK98}; for instance, it cannot be approximated
within any constant unless \P$=$\NP. However, these results do not
carry over to the $k$-cosparse projection problem, since the
objectives differ: In the optimization version of
\textsc{MinULR}$_0^=$, we wish to minimize the number $K$ of violated
equalities, while in \eqref{prob:ProjOpt}, the goal is minimizing the
distance of $\z$ to a given point (under the \emph{constraint} that
the number of nonzeros in $\Omeg\z$ does not exceed~$K$). Thus,
despite the link between the two problems exploited in the proof of
Theorem~\ref{thm:ProjNP}, (hypothetical) approximation guarantees for
this distance unfortunately do not yield any (non-)approximability
statements for~\eqref{prob:ProjOpt} by means of those
for~\textsc{MinULR}$_0^=$.

Thus, it remains a challenge to find (practically) efficient
approximation schemes for the $k$-cosparse projection
problem~\eqref{prob:ProjOpt}, or to establish further (perhaps
negative) results concerning its approximability. Other open questions
are the complexity of~\eqref{prob:ProjOpt} for $0<p\leq 1$, or
containment in {\NP} of~\eqref{prob:ProjOptDec} for
$p\in\N\setminus\{1,2\}$, cf. Remarks~\ref{rem:proofNotFor_pEQ1}
and~\ref{rem:NPcompleteness}.



%

\bibliographystyle{IEEEbib}
\bibliography{IEEEabrv}
\end{document}